\documentclass[english]{article}
\usepackage[T1]{fontenc}
\usepackage[latin9]{inputenc}
\usepackage{geometry}
\usepackage{color}
\usepackage{float}
\usepackage{booktabs}
\usepackage{slashed}
\usepackage{amstext}
\usepackage{amssymb}
\usepackage{graphicx}
\usepackage{setspace}

\makeatletter

\providecommand{\tabularnewline}{\\}

\usepackage{cite,times}
\date{}
\renewcommand{\textendash}{--}

\makeatother

\usepackage{babel}
\begin{document}

\title{Hierarchical Quantum Network using Hybrid Entanglement}

\author{Chitra Shukla$^{1,}$\thanks{Corresponding Author: shuklac@pcl.ac.cn; chitrashukla07@gmail.com},
Priya Malpani$^{2,}$\thanks{malpani.1@iitj.ac.in; }, Kishore Thapliyal$^{3,}$\thanks{kishore.thapliyal@upol.cz }\\
{\normalsize{}$^{1}$Center for Quantum Computing, Peng Cheng Laboratory,
Shenzhen 518055, People\textquoteright s Republic of China}\\
{\normalsize{}$^{2}$ Indian Institute of Technology Jodhpur, Jodhpur,
342037 India }\\
{\normalsize{}$^{3}$RCPTM, Joint Laboratory of Optics of Palacky
University and Institute of Physics of Academy of Science of }\\
{\normalsize{}the Czech Republic, Faculty of Science, Palacky University,
17. listopadu 12, 771 46, Olomouc, Czech Republic}}
\maketitle
\begin{abstract}
The advent of a new kind of entangled state known as hybrid entangled
state, i.e., entanglement between different degrees of freedom, makes
it possible to perform various quantum computational and communication
tasks with lesser amount of resources. Here, we aim to exploit the
advantage of these entangled states in communication over quantum
networks. Unfortunately, the entanglement shared over the network
deteriorates due to its unavoidable interaction with surroundings.
Thus, an entanglement concentration protocol is proposed to obtain
a maximally entangled hybrid Omega-type state from the corresponding
non-maximally entangled states. The advantage of the proposed entanglement
concentration protocol is that it is feasible to implement this protocol
with linear optical components and present technology. The corresponding
linear optical quantum circuit is provided for experimental realizations,
while the success probability of the concentration protocol is also
reported. Thereafter, we propose an application of maximally entangled
hybrid state in the hierarchical quantum teleportation network by
performing information splitting using Omega-type state, which is
also the first hierarchical quantum communication scheme in the hybrid
domain so far. The present hybrid entangled state has advantage in
circumventing Pauli operations on the coherent state by polarization
rotation of single qubit, which can be performed with lesser errors.
\end{abstract}

\section{Introduction\label{sec:Introduction} }

Quantum entanglement is an irreplaceable quantum resource that plays
a very significant role in the field of quantum information processing
\cite{Nielsen_Chuang}. It is an essential element for various applications
in the fields of quantum computation \cite{QC_CV_Cluster1} and communication
\cite{Nielsen_Chuang}, such as quantum teleportation \cite{tele},
dense coding \cite{dense_coding}, quantum key distribution \cite{ekert},
quantum key agreement \cite{QKA1,QKA2}, quantum secret sharing \cite{QSS},
quantum secure direct communication \cite{review,QSDC-Review}. In
the recent times, communication over quantum networks and internet
is of primary interest \cite{Q_Internet,End-to-end_network,Remote_creation},
which often uses entanglement and teleportation. Interestingly, quantum
entanglement is not sufficient for most of these applications, but
maximal quantum entanglement is necessary. For example, quantum teleportation
\cite{tele} and its variants \cite{Ba_An_Hierarchically_controlling_QT}
(and references therein) essentially require maximally entangled state
(MES) to achieve the unit success probability for deterministic communication.
Unavailability of such MES would lead to success probability less
than unity and thus the communication becomes probabilistic \cite{Probabilistic_Teleporation}
or the fidelity of the transmitted state not unity \cite{HJRSP}.
Similarly, security and efficiency of the secure communication schemes
require MESs \cite{Pathak_Book}. The practical problems associated
with exploiting MES is its fragile nature due to decoherence induced
by the environment. Specifically, while managing the storage, transmission,
and processing, the shared entangled states (quantum channel) interact
with the noisy environment which causes degradation in entanglement
quality and thus the performance of the quantum communication protocols.
Basically, it is impossible to avoid the decoherence which imposes
a limitation in the successful execution of quantum technology as
it reduces a MES into a non-MES or even a mixed states. Therefore,
it is recommended to transform a non-MES into a MES before performing
any quantum communication over a network. There are two important
techniques in the literature to recover MES. Namely, entanglement
concentration protocol (ECP) to recover a MES from pure non-MESs and
entanglement purification protocol (EPP) for mixed non-MESs. 

First such attempt to improve the quality of entanglement was performed
by Bennett et al. by introducing a Schmidt decomposition based ECP
\cite{ECP_Bennet} and an EPP \cite{EPP_Bennet}. These pioneering
works were followed by various ECPs \cite{ECP_Bose,ECP_Zhao,ECP_Sheng2,Our_ECP1,ECP_Sheng_2019,Wang_2019}
and EPPs \cite{EPP_Pan,EPP_yamamoto,EPP_Zwerger1,Sheng_EPP_Scin_Rep},
which were restricted to the discrete variables (DVs). Independently,
an extensive effort has been put in the study of continuous variable
(CV) entanglement in theory \cite{Barry_2} (and references therein)
and experiments \cite{Experimental_CV_Cluster,Experimental_CV_Cluster1},
with applications in various quantum cryptographic \cite{CV_QKD}
and information processing tasks \cite{CVQKD2,Nguan_Ba_An}. This
motivated CV ECPs \cite{W-ECS,Our_OECP} reported in the recent past.
Several experimental realizations and applications of ECPs and EPPs
play an important role in the development of quantum technology \cite{ECP_Experiment,EPP_Experiment}
and references therein. 

Traditionally, entanglement encoded in both DV (such as polarization
states of photons) and CV (superposition of coherent states with opposite
phases $|\pm\alpha\rangle$ and large value of amplitude $\alpha$)
has been adopted separately in the literature for various applications
in DV \cite{review,QSDC-Review,Pathak_Book} and CV \cite{CV_QKD,CVQKD2}
quantum communication, respectively. Both the approaches have their
own pros and cons. In 2006, a new DV-CV approach has been developed
\cite{First_hybrid_Quantum_Repeater} by combining the above two approaches
to harness the advantages of discrete and continuous degrees of freedom
(DOFs) together. This new DV-CV approach has been referred to as hybrid
entangled states (HESs) where the entanglement is generated between
different DOFs of a particle/mode pair. Hybrid entanglement based
quantum information processing tasks require fewer resources because
it can be correlated within a single particle and thus outperforms
the conventional method of using only a single type of DOF. In \cite{First_hybrid_Quantum_Repeater},
the first hybrid quantum repeater using bright coherent light was
proposed. Since then several researchers have made efforts independently
for HES generation, such as generation of HES of light \cite{Hybrid_Generation1},
HES between a single-photon polarization qubit and a coherent state
\cite{Hybrid_Generation2}, HES using interaction between DV and CV
states \cite{Ba_An_Hybrid_generation3}, HES between particle-like
and wave-like optical qubits induced by measurement \cite{Remote_creation}.
Interestingly, many different applications have been reported based
on HES, for example, near-deterministic quantum teleportation and
resource-efficient quantum computation \cite{Near-deterministic_teleportation},
hybrid long distance entanglement distribution protocol \cite{Hybrid_long-distance_distribution},
hybrid quantum information processing \cite{Hybrid _QIP}. Further,
quantum teleportation between DV and CV optical qubits have been proposed
\cite{QT_optical_qubit_1,QT_optical_qubit_2} in addition to remote
preparation of CV qubits using HES \cite{Remote_Preparation} and
more recently an efficient quantum teleportation based on DV-CV interaction
mechanism \cite{Efficient_QT}. Applications of HESs in remote state
preparation and quantum steering \cite{Hybrid_RSP_Steering} and a
remote preparation for single-photon two-qubit HES using hyperentanglement
\cite{Sci-Rep_RSP_hyper_hybrid_2019} are also proposed. Interestingly,
first EPP for quantum repeaters \cite{Hybrid_EPP_Sheng}, ECP assisted
with single coherent state \cite{HES_Cluster_ECP}, ECP and EPP in
spatially separated spins in nitrogen-vacancy centers \cite{ECP_EPP_2019}
for HESs are also proposed. Hence, HESs have already established to
be a very promising direction in advancing the quantum technology
because of the combined advantages of two or more DOFs. 

Therefore, inspired by these applications here we aim to propose communication
over a hierarchical quantum teleportation network using Omega-type
HES, which is referred to as a deterministic hybrid hierarchical quantum
information splitting (HQIS). However, while network communication,
Omega-type HES is expected to decohere due to the effect of environment,
hence for an efficient implementation of communication over network
first we have to propose an ECP for Omega-type HES which has the significance
of being entangled in both DV and CV. We further aim to report an
HQIS of the MES generated by the proposed ECP for HES in a hitherto
unexplored domain of quantum communication. Specifically, applications
of DV Omega state have been discussed in the past \cite{QD,HQIS}
and thus CV and hybrid (DV-CV) Omega states are also expected to be
useful in quantum communication tasks. Here, we have focused on quantum
communication based on maximally HES, and thus it becomes essential
to design an ECP for Omega-type HES to protect it from the influence
of noise. To the best of our knowledge, neither an ECP nor an application
in hierarchical communication for Omega-type HES has yet been proposed.
A hierarchical scheme, namely hierarchical quantum information splitting
\cite{HQIS}, can be viewed as a variant of multiparty teleportation,
where the receivers (agents) are graded in accordance with their power
for the reconstruction of an unknown quantum state sent by a sender
(boss). A couple of interesting proposals on hierarchical communication
schemes have been proposed in the recent years, for example, a generalized
structure of HQIS \cite{HQIS} using $(n+1)-$qubit states, integrated
hierarchical dynamic quantum secret sharing \cite{HDQSS} combining
features of hierarchy and dynamism, a unique scheme of hierarchical
joint remote state preparation \cite{HJRSP} with joint preparation
to protect the secrecy of a sensitive information, and hierarchically
controlled quantum teleportation \cite{Ba_An_Hierarchically_controlling_QT}.
All these hierarchical schemes are shown to have wide applications
in the real world scenario (\cite{Ba_An_Hierarchically_controlling_QT,HJRSP,HQIS,HDQSS}
and references therein). 

The rest of the paper is organized as follows. In Sec. \ref{sec:Omega-type-hybrid},
Omega-type HES is introduced. In Sec. \ref{sec:Entanglement-Concentration},
we have described our ECP for Omega-type HES from non-MESs using a
single- and two-mode coherent states with four parameters. Subsequently,
in Sec. \ref{sec:Success-probability}, the success probability for
ECP has been discussed. Furthermore, a deterministic hybrid HQIS scheme
using Omega-type HES has been proposed before concluding in Sec. \ref{sec:Conclusion}. 

\begin{figure}
\begin{centering}
\includegraphics[scale=0.4]{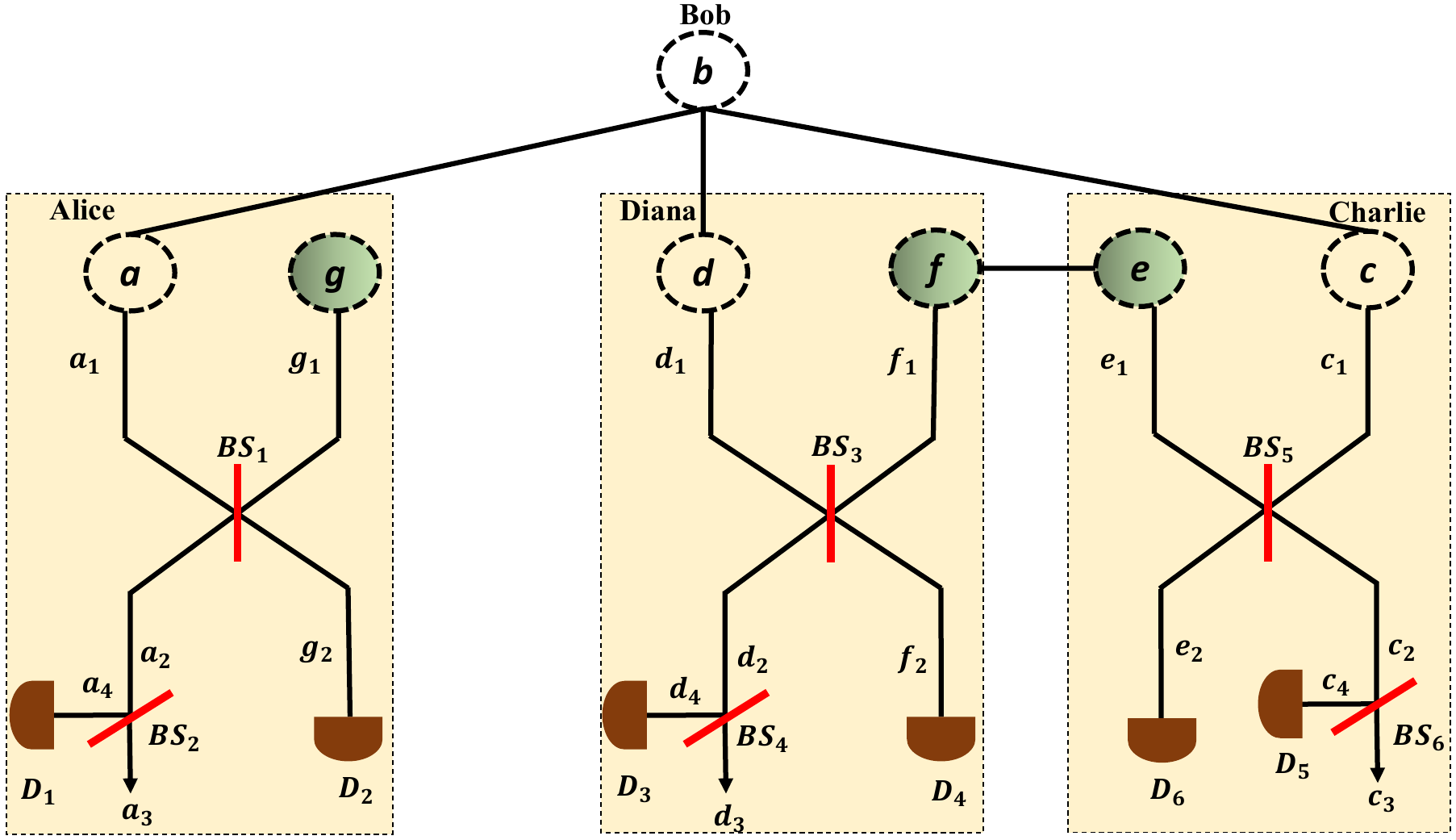}
\par\end{centering}
\caption{\label{fig:(Color-online)-The}(Color online) The schematic diagram
of the proposed ECP for Omega-type HES. Initially, Alice, Bob, Charlie,
and Diana share hybrid Omega-type non-MES. The subscripts $a,b,c$,
and $d$ denote the states with Alice, Bob, Charlie, and Diana, respectively.
Alice prepares a single-mode superposition of coherent state in the
spatial mode $g_{1}$, and Charlie (or Diana) prepares and shares
a two-mode entangled coherent state in the spatial mode $e_{1}$ and
$f_{1}$, respectively. Modes of the HES to be concentrated are mixed
with the auxiliary modes at the symmetric beamsplitters $BS_{1},$
$BS_{3}$, and $BS_{5}$. Further, to obtain the hybrid Omega-type
MES, beamsplitters $BS_{2}$, $BS_{4}$, and $BS_{6}$ are useful.
Spatial modes after evolution from an optical element are labeled
as $i_{j}\rightarrow i_{j+1}$. Further, $D_{x}$ are the photon detectors. }
\end{figure}

\section{Omega-type hybrid entangled state \label{sec:Omega-type-hybrid}}

An Omega-type hybrid MES can be written as

\begin{equation}
\begin{array}{ccc}
\left|\Omega\right\rangle _{abcd} & = & \frac{1}{2}\left(\left|0_{L}\right\rangle \left|0_{L}\right\rangle \left|0_{L}\right\rangle \left|0_{L}\right\rangle +\left|0_{L}\right\rangle \left|1_{L}\right\rangle \left|1_{L}\right\rangle \left|0_{L}\right\rangle +\left|1_{L}\right\rangle \left|0_{L}\right\rangle \left|0_{L}\right\rangle \left|1_{L}\right\rangle -\left|1_{L}\right\rangle \left|1_{L}\right\rangle \left|1_{L}\right\rangle \left|1_{L}\right\rangle \right),\end{array}\label{eq:omega logic}
\end{equation}
where an optical logical hybrid qubit $\left|0_{L}\right\rangle $
and $\left|1_{L}\right\rangle $ is defined in basis $\left\{ \left|0_{L}\right\rangle =\left|+\right\rangle \left|\alpha\right\rangle ,\left|1_{L}\right\rangle =\left|-\right\rangle \left|-\alpha\right\rangle \right\} $
with single qubit polarization states $\left|\pm\right\rangle =\frac{\left(\left|H\right\rangle \pm\left|V\right\rangle \right)}{\sqrt{2}}$
and coherent states $\left|\pm\alpha\right\rangle $. Here, the subscripts
$a,b,c$, and $d$ represent states possessed by Alice, Bob, Charlie,
and Diana, respectively. Here, a logical hybrid qubit encoded in polarization
and coherent state is analogous to Ref. \cite{Near-deterministic_teleportation,Omkar},
and the logical state can be identified as Omega state \cite{9_Family_of_states,Pati_Omega}.
Thus, Eq. (\ref{eq:omega logic}) can be expressed as

\begin{equation}
\begin{array}{lcl}
\left|\Omega\right\rangle _{abcd} & = & \frac{1}{2}\left(\left|+\right\rangle \left|+\right\rangle \left|+\right\rangle \left|+\right\rangle \left|\alpha\right\rangle \left|\alpha\right\rangle \left|\alpha\right\rangle \left|\alpha\right\rangle +\left|+\right\rangle \left|-\right\rangle \left|-\right\rangle \left|+\right\rangle \left|\alpha\right\rangle \left|-\alpha\right\rangle \left|-\alpha\right\rangle \left|\alpha\right\rangle \right.\\
 & + & \left.\left|-\right\rangle \left|+\right\rangle \left|+\right\rangle \left|-\right\rangle \left|-\alpha\right\rangle \left|\alpha\right\rangle \left|\alpha\right\rangle \left|-\alpha\right\rangle -\left|-\right\rangle \left|-\right\rangle \left|-\right\rangle \left|-\right\rangle \left|-\alpha\right\rangle \left|-\alpha\right\rangle \left|-\alpha\right\rangle \left|-\alpha\right\rangle \right).
\end{array}\label{eq:omega}
\end{equation}
As we have already discussed that due to the effect of the surroundings
the hybrid MES in Eq. (\ref{eq:omega}) transforms into a non-MES.
Specifically, during the sharing of entanglement after its local preparation
in the practical scenario, the MES becomes partially-entangled. The
hybrid non-MES corresponding to Eq. (\ref{eq:omega}) can be written
as

\begin{equation}
\begin{array}{lcl}
\left|\Omega^{\prime}\right\rangle _{abcd} & = & N_{0}\left(\zeta\left|+\right\rangle \left|+\right\rangle \left|+\right\rangle \left|+\right\rangle \left|\alpha\right\rangle \left|\alpha\right\rangle \left|\alpha\right\rangle \left|\alpha\right\rangle +\beta\left|+\right\rangle \left|-\right\rangle \left|-\right\rangle \left|+\right\rangle \left|\alpha\right\rangle \left|-\alpha\right\rangle \left|-\alpha\right\rangle \left|\alpha\right\rangle \right.\\
 & + & \left.\gamma\left|-\right\rangle \left|+\right\rangle \left|+\right\rangle \left|-\right\rangle \left|-\alpha\right\rangle \left|\alpha\right\rangle \left|\alpha\right\rangle \left|-\alpha\right\rangle -\delta\left|-\right\rangle \left|-\right\rangle \left|-\right\rangle \left|-\right\rangle \left|-\alpha\right\rangle \left|-\alpha\right\rangle \left|-\alpha\right\rangle \left|-\alpha\right\rangle \right),
\end{array}\label{eq:omega less entangled}
\end{equation}
where the normalization constant $N_{0}=\left[\zeta^{2}+\beta^{2}+\gamma^{2}+\delta^{2}\right]^{-\frac{1}{2}}$
is unity. Probability amplitudes $\zeta,\beta,\gamma,$ and $\delta$
are non-zero and considered real. Here, all probability amplitudes
are not same because that will correspond to MES (\ref{eq:omega}).
We aim to obtain a hybrid Omega-type MES (\ref{eq:omega}) from Eq.
(\ref{eq:omega less entangled}) by performing ECP proposed in the
following section.

Before we proceed further, it is imperative to discuss the 
preparation of quantum states in Eqs.~(\ref{eq:omega})-(\ref{eq:omega less entangled}). For generation of hybrid Omega-type MES, we require two single photon pairs and a hybrid entangled state of two logical qubits $\left|\psi_{H2}\right\rangle$. Specifically, this particular hybrid entangled state can be prepared using two hybrid pairs in state $\left|\psi_{H1}\right\rangle$ and a single photon pair in the Bell state $\left|\phi^{+}\right\rangle =\frac{\left|HH\right\rangle +\left|VV\right\rangle }{\sqrt{2}}$. However, if all these resource states are non-maximally entangled, then the two logical qubit state $\left|\psi_{H2}\right\rangle$ will be obtained from
\begin{equation}
\begin{array}{lcl}
\left|\psi^\prime_{H2}\right\rangle &=& \left|\psi_{H1}\right\rangle_1\otimes \left|\phi^{+}\right\rangle \otimes \left|\psi_{H1}\right\rangle_2
\\
&=&\left(a_0\left|H,\alpha\right\rangle +a_1\left|V,-\alpha\right\rangle \right)\left(b_0\left|HH\right\rangle +b_1\left|VV\right\rangle \right)\left(c_0\left|H,\alpha\right\rangle +c_1\left|V,-\alpha\right\rangle \right)
\end{array}
\end{equation}
by performing a parity check measurement on the single photon of $\left|\psi_{H1}\right\rangle_1$ ($\left|\psi_{H1}\right\rangle_2$) and the first qubit of the Bell state after application of a Hadamard operation (the second qubit of the Bell state). Here, $\sum_j A_j^2=1\,\forall\, A\in \{a,b,c \} $ and Hadamard operation on polarization qubits correspond to the use of a half-wave plate at 22.5$^{\rm o}$. Conditioned on the the same parity qubits in both parity check measurements, we finally perform quantum non-demolition measurement on both the qubits of the Bell state, i.e., measure both the qubits in $\{\left|+\right\rangle,\left|-\right\rangle \}$ basis. Thus, final state in the superposition of two logical qubit states after an application of Hadamard operation on both the single photon states becomes
\begin{equation}
\begin{array}{lcl}
\left|\psi_{H2}\right\rangle &=& \beta_0\left|0_{L}\right\rangle \left|0_{L}\right\rangle+\beta_1 \left|0_{L}\right\rangle \left|1_{L}\right\rangle +\beta_2 \left|1_{L}\right\rangle \left|0_{L}\right\rangle -\beta_3\left|1_{L}\right\rangle \left|1_{L}\right\rangle,
\end{array}\label{eq:psiH2}
\end{equation}
where the probability amplitudes depend on the parameters of the initial states, such as $\beta_0=a_0b_0c_0$. A close look at Eqs.~(\ref{eq:omega logic}) and (\ref{eq:psiH2}) reveals us that we have to perform a CNOT operation with control on both these logical qubits and target on ancillae prepared in $\left|0_{L}\right\rangle$. 

It can also be accomplished using $\left|\psi_{H2}\right\rangle$ with coherent amplitude $\sqrt{2}\alpha$ and two Bell states 
\begin{equation}
\begin{array}{lcl}
\left|\psi^\prime_{H4}\right\rangle &=&\left|\phi^{+}\right\rangle_1 \otimes  \left|\psi_{H2}\right\rangle\otimes \left|\phi^{+}\right\rangle_2.
\end{array}
\end{equation}
After performing a Hadamard operation on one qubit each of the Bell states and both single photons in $\left|\psi_{H2}\right\rangle$ we can perform parity check measurement on one qubit of the first (second) Bell state and the first (second) single photon in $\left|\psi_{H2}\right\rangle$. Further, we condition the same parity outcomes on both these measurements and performing quantum non-demolition measurement on both the single photons of $\left|\psi_{H2}\right\rangle$. Finally, we perform a Hadamard operation on all the single photons and pass both the coherent modes through respective beamsplitters to obtain the desired state (\ref{eq:omega less entangled}).
Note that a beamsplitter $(BS)$ transforms the two input coherent states $|\alpha\rangle$
and $|\beta\rangle$ as 
\begin{equation}
BS|\alpha\rangle|\beta\rangle\rightarrow\left|\frac{\alpha+\beta}{\sqrt{2}}\right\rangle \left|\frac{\alpha-\beta}{\sqrt{2}}\right\rangle.\label{eq:BS_Function}
\end{equation}
Interestingly, generation of hybrid Omega-type MES (\ref{eq:omega}) requires that all three Bell states and two hybrid pairs should be maximally entangled. If any of these entangled states is not maximally entangled the generated quantum state will be of the form (\ref{eq:omega less entangled}) and require entanglement concentration proposed here.
The hybrid Omega-type MES (or non-MES) can be generated using different set of operations and input states, for instance using parity measurement on coherent states
\cite{Near-deterministic_teleportation} or Bell-like hybrid measurement on two three-qubit hybrid cluster states \cite{Omkar}.

\section{Entanglement concentration protocol for Omega-type hybrid entangled
state \label{sec:Entanglement-Concentration}}

To begin with, first we propose an entanglement concentration for
Omega-type HES using a single- and a two-mode superpositions of coherent
states. To obtain a MES of hybrid Omega-type, we assume that Alice,
Bob, Charlie, and Diana share the hybrid Omega-type non-MES as described
in Eq. (\ref{eq:omega less entangled}). The optical circuit of ECP
for Omega-type HES is shown in Fig. \ref{fig:(Color-online)-The}.
The subscripts $a,$$b,$$c$, and $d$ represent to the spatial modes
of Alice, Bob, Charlie, and Diana, respectively. Further, Charlie
(or equivalently Diana) prepares and shares a superposition of two-mode
coherent states $\left|\psi\right\rangle _{e_{1}f_{1}}$ in the spatial
mode $e_{1}$ and $f_{1}$ among Charlie and Diana, respectively, 

\begin{equation}
\left|\psi\right\rangle _{e_{1}f_{1}}=N_{1}\left(\zeta\left|-\alpha\right\rangle \left|\alpha\right\rangle +\beta\left|\alpha\right\rangle \left|\alpha\right\rangle +\gamma\left|-\alpha\right\rangle \left|-\alpha\right\rangle +\delta\left|\alpha\right\rangle \left|-\alpha\right\rangle \right),\label{eq:2_mode_ancilla}
\end{equation}
where $N_{1}=\left[\zeta^{2}+\beta^{2}+\delta{}^{2}+\gamma^{2}+2(\zeta\beta+\zeta\gamma+\delta\beta+\gamma\delta)e^{-2|\alpha|^{2}}+2(\gamma\beta+\delta\zeta)e^{-4|\alpha|^{2}}\right]^{-\frac{1}{2}}$
is the normalization constant. Therefore, the combined state of the
system can be expressed as 

\begin{equation}
\left|\phi\right\rangle _{abcde_{1}f_{1}}=\left|\Omega^{\prime}\right\rangle _{abcd}\otimes\left|\psi\right\rangle _{e_{1}f_{1}}.\label{eq:combined}
\end{equation}
Here and in what follows, we label the spatial modes on which we perform
operations as $a_{1},$ $c_{1}$ and $d_{1}$ while $b$ mode is unchanged
as we do not perform any operation on this mode. Further, all the
operations are only performed on the CV DOF and thus polarization
remains unchanged. Thus, we can write

\begin{equation}
\begin{array}{ll}
\left|\phi\right\rangle _{a_{1}bc_{1}d_{1}e_{1}f_{1}} & =N_{1}\left(\zeta^{2}\left|+\right\rangle \left|+\right\rangle \left|+\right\rangle \left|+\right\rangle \left|\alpha\right\rangle \left|\alpha\right\rangle \left|\alpha\right\rangle \left|\alpha\right\rangle \left|-\alpha\right\rangle \left|\alpha\right\rangle +\zeta\beta\left|+\right\rangle \left|-\right\rangle \left|-\right\rangle \left|+\right\rangle \left|\alpha\right\rangle \left|-\alpha\right\rangle \left|-\alpha\right\rangle \left|\alpha\right\rangle \left|-\alpha\right\rangle \left|\alpha\right\rangle \right.\\
 & +\gamma\zeta\left|-\right\rangle \left|+\right\rangle \left|+\right\rangle \left|-\right\rangle \left|-\alpha\right\rangle \left|\alpha\right\rangle \left|\alpha\right\rangle \left|-\alpha\right\rangle \left|-\alpha\right\rangle \left|\alpha\right\rangle -\zeta\delta\left|-\right\rangle \left|-\right\rangle \left|-\right\rangle \left|-\right\rangle \left|-\alpha\right\rangle \left|-\alpha\right\rangle \left|-\alpha\right\rangle \left|-\alpha\right\rangle \left|-\alpha\right\rangle \left|\alpha\right\rangle \\
 & +\beta\zeta\left|+\right\rangle \left|+\right\rangle \left|+\right\rangle \left|+\right\rangle \left|\alpha\right\rangle \left|\alpha\right\rangle \left|\alpha\right\rangle \left|\alpha\right\rangle \left|\alpha\right\rangle \left|\alpha\right\rangle +\beta^{2}\left|+\right\rangle \left|-\right\rangle \left|-\right\rangle \left|+\right\rangle \left|\alpha\right\rangle \left|-\alpha\right\rangle \left|-\alpha\right\rangle \left|\alpha\right\rangle \left|\alpha\right\rangle \left|\alpha\right\rangle \\
 & +\beta\gamma\left|-\right\rangle \left|+\right\rangle \left|+\right\rangle \left|-\right\rangle \left|-\alpha\right\rangle \left|\alpha\right\rangle \left|\alpha\right\rangle \left|-\alpha\right\rangle \left|\alpha\right\rangle \left|\alpha\right\rangle -\delta\beta\left|-\right\rangle \left|-\right\rangle \left|-\right\rangle \left|-\right\rangle \left|-\alpha\right\rangle \left|-\alpha\right\rangle \left|-\alpha\right\rangle \left|-\alpha\right\rangle \left|\alpha\right\rangle \left|\alpha\right\rangle \\
 & +\gamma\zeta\left|+\right\rangle \left|+\right\rangle \left|+\right\rangle \left|+\right\rangle \left|\alpha\right\rangle \left|\alpha\right\rangle \left|\alpha\right\rangle \left|\alpha\right\rangle \left|-\alpha\right\rangle \left|-\alpha\right\rangle +\beta\gamma\left|+\right\rangle \left|-\right\rangle \left|-\right\rangle \left|+\right\rangle \left|\alpha\right\rangle \left|-\alpha\right\rangle \left|-\alpha\right\rangle \left|\alpha\right\rangle \left|-\alpha\right\rangle \left|-\alpha\right\rangle \\
 & +\gamma^{2}\left|-\right\rangle \left|+\right\rangle \left|+\right\rangle \left|-\right\rangle \left|-\alpha\right\rangle \left|\alpha\right\rangle \left|\alpha\right\rangle \left|-\alpha\right\rangle \left|-\alpha\right\rangle \left|-\alpha\right\rangle -\delta\gamma\left|-\right\rangle \left|-\right\rangle \left|-\right\rangle \left|-\right\rangle \left|-\alpha\right\rangle \left|-\alpha\right\rangle \left|-\alpha\right\rangle \left|-\alpha\right\rangle \left|-\alpha\right\rangle \left|-\alpha\right\rangle \\
 & +\zeta\delta\left|+\right\rangle \left|+\right\rangle \left|+\right\rangle \left|+\right\rangle \left|\alpha\right\rangle \left|\alpha\right\rangle \left|\alpha\right\rangle \left|\alpha\right\rangle \left|\alpha\right\rangle \left|-\alpha\right\rangle +\beta\delta\left|+\right\rangle \left|-\right\rangle \left|-\right\rangle \left|+\right\rangle \left|\alpha\right\rangle \left|-\alpha\right\rangle \left|-\alpha\right\rangle \left|\alpha\right\rangle \left|\alpha\right\rangle \left|-\alpha\right\rangle \\
 & +\left.\delta\gamma\left|-\right\rangle \left|+\right\rangle \left|+\right\rangle \left|-\right\rangle \left|-\alpha\right\rangle \left|\alpha\right\rangle \left|\alpha\right\rangle \left|-\alpha\right\rangle \left|\alpha\right\rangle \left|-\alpha\right\rangle -\delta^{2}\left|-\right\rangle \left|-\right\rangle \left|-\right\rangle \left|-\right\rangle \left|-\alpha\right\rangle \left|-\alpha\right\rangle \left|-\alpha\right\rangle \left|-\alpha\right\rangle \left|\alpha\right\rangle \left|-\alpha\right\rangle \right).
\end{array}\label{eq:combind_state}
\end{equation}

First of all, Charlie (Diana) passes modes $c_{1}$ and $e_{1}$ $(d_{1}$
and $f_{1})$ through 50:50 (symmetric or balanced) beamsplitter $BS_{5},\,\,(BS_{3})$.
Using
Eq. (\ref{eq:BS_Function}) the state post-beamsplitter can be written
as
\begin{equation}
\begin{array}{l}
\left|\phi\right\rangle _{a_{1}bc_{2}d_{2}e_{2}f_{2}}\\
=N_{1}\left(\zeta^{2}\left|+\right\rangle \left|+\right\rangle \left|+\right\rangle \left|+\right\rangle \left|\alpha\right\rangle \left|\alpha\right\rangle \left|0\right\rangle \left|\sqrt{2}\alpha\right\rangle \left|\sqrt{2}\alpha\right\rangle \left|0\right\rangle +\zeta\beta\left|+\right\rangle \left|-\right\rangle \left|-\right\rangle \left|+\right\rangle \left|\alpha\right\rangle \left|-\alpha\right\rangle \left|-\sqrt{2}\alpha\right\rangle \left|\sqrt{2}\alpha\right\rangle \left|0\right\rangle \left|0\right\rangle \right.\\
+\gamma\zeta\left|-\right\rangle \left|+\right\rangle \left|+\right\rangle \left|-\right\rangle \left|-\alpha\right\rangle \left|\alpha\right\rangle \left|0\right\rangle \left|0\right\rangle \left|\sqrt{2}\alpha\right\rangle \left|-\sqrt{2}\alpha\right\rangle -\zeta\delta\left|-\right\rangle \left|-\right\rangle \left|-\right\rangle \left|-\right\rangle \left|-\alpha\right\rangle \left|-\alpha\right\rangle \left|-\sqrt{2}\alpha\right\rangle \left|0\right\rangle \left|0\right\rangle \left|-\sqrt{2}\alpha\right\rangle \\
+\beta\zeta\left|+\right\rangle \left|+\right\rangle \left|+\right\rangle \left|+\right\rangle \left|\alpha\right\rangle \left|\alpha\right\rangle \left|\sqrt{2}\alpha\right\rangle \left|\sqrt{2}\alpha\right\rangle \left|0\right\rangle \left|0\right\rangle +\beta^{2}\left|+\right\rangle \left|-\right\rangle \left|-\right\rangle \left|+\right\rangle \left|\alpha\right\rangle \left|-\alpha\right\rangle \left|0\right\rangle \left|\sqrt{2}\alpha\right\rangle \left|-\sqrt{2}\alpha\right\rangle \left|0\right\rangle \\
+\beta\gamma\left|-\right\rangle \left|+\right\rangle \left|+\right\rangle \left|-\right\rangle \left|-\alpha\right\rangle \left|\alpha\right\rangle \left|\sqrt{2}\alpha\right\rangle \left|0\right\rangle \left|0\right\rangle \left|-\sqrt{2}\alpha\right\rangle -\delta\beta\left|-\right\rangle \left|-\right\rangle \left|-\right\rangle \left|-\right\rangle \left|-\alpha\right\rangle \left|-\alpha\right\rangle \left|0\right\rangle \left|0\right\rangle \left|-\sqrt{2}\alpha\right\rangle \left|-\sqrt{2}\alpha\right\rangle \\
+\gamma\zeta\left|+\right\rangle \left|+\right\rangle \left|+\right\rangle \left|+\right\rangle \left|\alpha\right\rangle \left|\alpha\right\rangle \left|0\right\rangle \left|0\right\rangle \left|\sqrt{2}\alpha\right\rangle \left|\sqrt{2}\alpha\right\rangle +\beta\gamma\left|+\right\rangle \left|-\right\rangle \left|-\right\rangle \left|+\right\rangle \left|\alpha\right\rangle \left|-\alpha\right\rangle \left|-\sqrt{2}\alpha\right\rangle \left|0\right\rangle \left|0\right\rangle \left|\sqrt{2}\alpha\right\rangle \\
+\gamma^{2}\left|-\right\rangle \left|+\right\rangle \left|+\right\rangle \left|-\right\rangle \left|-\alpha\right\rangle \left|\alpha\right\rangle \left|0\right\rangle \left|\sqrt{2}\alpha\right\rangle \left|\sqrt{2}\alpha\right\rangle \left|0\right\rangle -\delta\gamma\left|-\right\rangle \left|-\right\rangle \left|-\right\rangle \left|-\right\rangle \left|-\alpha\right\rangle \left|-\alpha\right\rangle \left|-\sqrt{2}\alpha\right\rangle \left|-\sqrt{2}\alpha\right\rangle \left|0\right\rangle \left|0\right\rangle \\
+\zeta\delta\left|+\right\rangle \left|+\right\rangle \left|+\right\rangle \left|+\right\rangle \left|\alpha\right\rangle \left|\alpha\right\rangle \left|\sqrt{2}\alpha\right\rangle \left|0\right\rangle \left|0\right\rangle \left|\sqrt{2}\alpha\right\rangle +\beta\delta\left|+\right\rangle \left|-\right\rangle \left|-\right\rangle \left|+\right\rangle \left|\alpha\right\rangle \left|-\alpha\right\rangle \left|0\right\rangle \left|0\right\rangle \left|-\sqrt{2}\alpha\right\rangle \left|\sqrt{2}\alpha\right\rangle \\
+\left.\delta\gamma\left|-\right\rangle \left|+\right\rangle \left|+\right\rangle \left|-\right\rangle \left|-\alpha\right\rangle \left|\alpha\right\rangle \left|\sqrt{2}\alpha\right\rangle \left|-\sqrt{2}\alpha\right\rangle \left|0\right\rangle \left|0\right\rangle -\delta^{2}\left|-\right\rangle \left|-\right\rangle \left|-\right\rangle \left|-\right\rangle \left|-\alpha\right\rangle \left|-\alpha\right\rangle \left|0\right\rangle \left|-\sqrt{2}\alpha\right\rangle \left|-\sqrt{2}\alpha\right\rangle \left|0\right\rangle \right).
\end{array}\label{eq:}
\end{equation}
Postselecting the condition that there is no photon in modes $e_{2}$
and $f_{2}$, the state in these cases can be written as 

\begin{equation}
\begin{array}{ll}
\left|\phi\right\rangle _{a_{1}bc_{2}d_{2}} & =N_{1}\left(\zeta\beta\left|+\right\rangle \left|-\right\rangle \left|-\right\rangle \left|+\right\rangle \left|\alpha\right\rangle \left|-\alpha\right\rangle \left|-\sqrt{2}\alpha\right\rangle \left|\sqrt{2}\alpha\right\rangle +\beta\zeta\left|+\right\rangle \left|+\right\rangle \left|+\right\rangle \left|+\right\rangle \left|\alpha\right\rangle \left|\alpha\right\rangle \left|\sqrt{2}\alpha\right\rangle \left|\sqrt{2}\alpha\right\rangle \right.\\
 & -\left.\delta\gamma\left|-\right\rangle \left|-\right\rangle \left|-\right\rangle \left|-\right\rangle \left|-\alpha\right\rangle \left|-\alpha\right\rangle \left|-\sqrt{2}\alpha\right\rangle \left|-\sqrt{2}\alpha\right\rangle +\delta\gamma\left|-\right\rangle \left|+\right\rangle \left|+\right\rangle \left|-\right\rangle \left|-\alpha\right\rangle \left|\alpha\right\rangle \left|\sqrt{2}\alpha\right\rangle \left|-\sqrt{2}\alpha\right\rangle \right).
\end{array}\label{eq:-1}
\end{equation}
Furthermore, Charlie and Diana let modes $c_{2}$ and $d_{2}$ pass
through 50:50 beamsplitters $BS_{6}$ and $BS_{4}$, respectively,
and using Eq. (\ref{eq:BS_Function}) the state becomes 

\begin{equation}
\begin{array}{ll}
\left|\phi\right\rangle _{a_{1}bc_{3}c_{4}d_{3}d_{4}} & =N_{1}\left(\zeta\beta\left|+\right\rangle \left|-\right\rangle \left|-\right\rangle \left|+\right\rangle \left|\alpha\right\rangle \left|-\alpha\right\rangle \left|-\alpha\right\rangle \left|-\alpha\right\rangle \left|\alpha\right\rangle \left|\alpha\right\rangle +\beta\zeta\left|+\right\rangle \left|+\right\rangle \left|+\right\rangle \left|+\right\rangle \left|\alpha\right\rangle \left|\alpha\right\rangle \left|\alpha\right\rangle \left|\alpha\right\rangle \left|\alpha\right\rangle \left|\alpha\right\rangle \right.\\
 & -\left.\delta\gamma\left|-\right\rangle \left|-\right\rangle \left|-\right\rangle \left|-\right\rangle \left|-\alpha\right\rangle \left|-\alpha\right\rangle \left|-\alpha\right\rangle \left|-\alpha\right\rangle \left|-\alpha\right\rangle \left|-\alpha\right\rangle +\delta\gamma\left|-\right\rangle \left|+\right\rangle \left|+\right\rangle \left|-\right\rangle \left|-\alpha\right\rangle \left|\alpha\right\rangle \left|\alpha\right\rangle \left|\alpha\right\rangle \left|-\alpha\right\rangle \left|-\alpha\right\rangle \right).
\end{array}\label{eq:-2}
\end{equation}
Modes $c_{4}$ and $d_{4}$ are redundant after this step, but a measurement
here would collapse the superposition. Thus, by performing a photon
number measurement on these two modes Charlie and Diana cannot distinguish
$|\pm\alpha\rangle$ as they have same photon number distribution.
Afterwards, the state can be written as 

\begin{equation}
\begin{array}{ll}
\left|\phi\right\rangle _{a_{1}bc_{3}d_{3}} & =N_{1}\left(\zeta\beta\left|+\right\rangle \left|-\right\rangle \left|-\right\rangle \left|+\right\rangle \left|\alpha\right\rangle \left|-\alpha\right\rangle \left|-\alpha\right\rangle \left|\alpha\right\rangle +\beta\zeta\left|+\right\rangle \left|+\right\rangle \left|+\right\rangle \left|+\right\rangle \left|\alpha\right\rangle \left|\alpha\right\rangle \left|\alpha\right\rangle \left|\alpha\right\rangle \right.\\
 & -\left.\delta\gamma\left|-\right\rangle \left|-\right\rangle \left|-\right\rangle \left|-\right\rangle \left|-\alpha\right\rangle \left|-\alpha\right\rangle \left|-\alpha\right\rangle \left|-\alpha\right\rangle +\delta\gamma\left|-\right\rangle \left|+\right\rangle \left|+\right\rangle \left|-\right\rangle \left|-\alpha\right\rangle \left|\alpha\right\rangle \left|\alpha\right\rangle \left|-\alpha\right\rangle \right).
\end{array}\label{eq:-3}
\end{equation}
Subsequently, Alice prepares an single-mode ancilla in superposition
of coherent states in spatial mode $g_{1}$ as 
\begin{equation}
\begin{array}{ccc}
\left|\phi\right\rangle _{g_{1}} & = & N_{2}\left(\zeta\beta\left|-\alpha\right\rangle +\delta\gamma\left|\alpha\right\rangle \right),\end{array}\label{eq:single-mode-coherent-state}
\end{equation}
where the normalization constant $N_{2}=\left[\zeta^{2}\beta^{2}+\delta{}^{2}\gamma^{2}+2\zeta\beta\delta\gamma e^{-2|\alpha|^{2}}\right]^{-\frac{1}{2}}$.
Thus, the combined state of the system can be expressed as 

\begin{equation}
\begin{array}{ll}
\left|\phi\right\rangle _{a_{1}bc_{3}d_{3}g_{1}} & =\left|\phi\right\rangle _{a_{1}bc_{3}d_{3}}\otimes\left|\phi\right\rangle _{g_{1}}\\
{\color{red}} & =N_{1}N_{2}\left(\zeta^{2}\beta^{2}\left|+\right\rangle \left|-\right\rangle \left|-\right\rangle \left|+\right\rangle \left|\alpha\right\rangle \left|-\alpha\right\rangle \left|-\alpha\right\rangle \left|\alpha\right\rangle \left|-\alpha\right\rangle +\zeta^{2}\beta^{2}\left|+\right\rangle \left|+\right\rangle \left|+\right\rangle \left|+\right\rangle \left|\alpha\right\rangle \left|\alpha\right\rangle \left|\alpha\right\rangle \left|\alpha\right\rangle \left|-\alpha\right\rangle \right.\\
 & +\zeta\beta\delta\gamma\left|+\right\rangle \left|-\right\rangle \left|-\right\rangle \left|+\right\rangle \left|\alpha\right\rangle \left|-\alpha\right\rangle \left|-\alpha\right\rangle \left|\alpha\right\rangle \left|\alpha\right\rangle +\zeta\beta\delta\gamma\left|+\right\rangle \left|+\right\rangle \left|+\right\rangle \left|+\right\rangle \left|\alpha\right\rangle \left|\alpha\right\rangle \left|\alpha\right\rangle \left|\alpha\right\rangle \left|\alpha\right\rangle \\
 & +\zeta\beta\delta\gamma\left|-\right\rangle \left|+\right\rangle \left|+\right\rangle \left|-\right\rangle \left|-\alpha\right\rangle \left|\alpha\right\rangle \left|\alpha\right\rangle \left|-\alpha\right\rangle \left|-\alpha\right\rangle -\zeta\beta\delta\gamma\left|-\right\rangle \left|-\right\rangle \left|-\right\rangle \left|-\right\rangle \left|-\alpha\right\rangle \left|-\alpha\right\rangle \left|-\alpha\right\rangle \left|-\alpha\right\rangle \left|-\alpha\right\rangle \\
 & +\left.\delta^{2}\gamma^{2}\left|-\right\rangle \left|+\right\rangle \left|+\right\rangle \left|-\right\rangle \left|-\alpha\right\rangle \left|\alpha\right\rangle \left|\alpha\right\rangle \left|-\alpha\right\rangle \left|\alpha\right\rangle -\delta^{2}\gamma^{2}\left|-\right\rangle \left|-\right\rangle \left|-\right\rangle \left|-\right\rangle \left|-\alpha\right\rangle \left|-\alpha\right\rangle \left|-\alpha\right\rangle \left|-\alpha\right\rangle \left|\alpha\right\rangle \right).
\end{array}\label{eq:-4}
\end{equation}
Alice further passes modes $a_{1}$ and $g_{1}$ through 50:50 beamsplitter
$BS_{1}$. Now, using Eq. (\ref{eq:BS_Function}) the post-beamsplitter
state can be expressed as 

\begin{equation}
\begin{array}{ll}
\left|\phi\right\rangle _{a_{2}bc_{3}d_{3}g_{2}} & =N_{1}N_{2}\left(\zeta^{2}\beta^{2}\left|+\right\rangle \left|-\right\rangle \left|-\right\rangle \left|+\right\rangle \left|0\right\rangle \left|-\alpha\right\rangle \left|-\alpha\right\rangle \left|\alpha\right\rangle \left|\sqrt{2}\alpha\right\rangle +\zeta^{2}\beta^{2}\left|+\right\rangle \left|+\right\rangle \left|+\right\rangle \left|+\right\rangle \left|0\right\rangle \left|\alpha\right\rangle \left|\alpha\right\rangle \left|\alpha\right\rangle \left|\sqrt{2}\alpha\right\rangle \right.\\
 & +\zeta\beta\delta\gamma\left|+\right\rangle \left|-\right\rangle \left|-\right\rangle \left|+\right\rangle \left|\sqrt{2}\alpha\right\rangle \left|-\alpha\right\rangle \left|-\alpha\right\rangle \left|\alpha\right\rangle \left|0\right\rangle +\zeta\beta\delta\gamma\left|+\right\rangle \left|+\right\rangle \left|+\right\rangle \left|+\right\rangle \left|\sqrt{2}\alpha\right\rangle \left|\alpha\right\rangle \left|\alpha\right\rangle \left|\alpha\right\rangle \left|0\right\rangle \\
 & +\zeta\beta\delta\gamma\left|-\right\rangle \left|+\right\rangle \left|+\right\rangle \left|-\right\rangle \left|-\sqrt{2}\alpha\right\rangle \left|\alpha\right\rangle \left|\alpha\right\rangle \left|-\alpha\right\rangle \left|0\right\rangle -\zeta\beta\delta\gamma\left|-\right\rangle \left|-\right\rangle \left|-\right\rangle \left|-\right\rangle \left|-\sqrt{2}\alpha\right\rangle \left|-\alpha\right\rangle \left|-\alpha\right\rangle \left|-\alpha\right\rangle \left|0\right\rangle \\
 & +\left.\delta^{2}\gamma^{2}\left|-\right\rangle \left|+\right\rangle \left|+\right\rangle \left|-\right\rangle \left|0\right\rangle \left|\alpha\right\rangle \left|\alpha\right\rangle \left|-\alpha\right\rangle \left|-\sqrt{2}\alpha\right\rangle -\delta^{2}\gamma^{2}\left|-\right\rangle \left|-\right\rangle \left|-\right\rangle \left|-\right\rangle \left|0\right\rangle \left|-\alpha\right\rangle \left|-\alpha\right\rangle \left|-\alpha\right\rangle \left|-\sqrt{2}\alpha\right\rangle \right).
\end{array}\label{eq:-5}
\end{equation}
She further postselects cases with no photon in mode $g_{2}$, and
that leads to a reduced state which can be written as

\begin{equation}
\begin{array}{ll}
\left|\phi\right\rangle _{a_{2}bc_{3}d_{3}} & =N_{1}N_{2}\zeta\beta\delta\gamma\left(\left|+\right\rangle \left|-\right\rangle \left|-\right\rangle \left|+\right\rangle \left|\sqrt{2}\alpha\right\rangle \left|-\alpha\right\rangle \left|-\alpha\right\rangle \left|\alpha\right\rangle +\left|+\right\rangle \left|+\right\rangle \left|+\right\rangle \left|+\right\rangle \left|\sqrt{2}\alpha\right\rangle \left|\alpha\right\rangle \left|\alpha\right\rangle \left|\alpha\right\rangle \right.\\
 & +\left.\left|-\right\rangle \left|+\right\rangle \left|+\right\rangle \left|-\right\rangle \left|-\sqrt{2}\alpha\right\rangle \left|\alpha\right\rangle \left|\alpha\right\rangle \left|-\alpha\right\rangle -\left|-\right\rangle \left|-\right\rangle \left|-\right\rangle \left|-\right\rangle \left|-\sqrt{2}\alpha\right\rangle \left|-\alpha\right\rangle \left|-\alpha\right\rangle \left|-\alpha\right\rangle \right).
\end{array}\label{eq:-6}
\end{equation}
Notice that the amplitude of coherent state in mode $a_{2}$ in Eq.
(\ref{eq:-6}) is amplified by a factor of $\sqrt{2}$ (cf. Eq. (\ref{eq:omega})),
which may have advantages in long-range quantum communication. However,
for the sake of accomplishment of ECP task here Alice passes mode
$a_{2}$ through a 50:50 beamsplitter $BS_{2}$ in order to concentrate
the desired MES (\ref{eq:omega}). Similar beamsplitter operations
were used on modes $c_{2}$ and $b_{2}$ by using $BS_{6}$ and $BS_{4}$
in Eq. (\ref{eq:-1}). Therefore, after the beamsplitter operation
on $a_{2}$ mode using Eq. (\ref{eq:BS_Function}), the evolved state
can be expressed as

\begin{equation}
\begin{array}{ll}
\left|\phi\right\rangle _{a_{3}a_{4}bc_{3}d_{3}} & =N_{1}N_{2}\zeta\beta\delta\gamma\left(\left|+\right\rangle \left|-\right\rangle \left|-\right\rangle \left|+\right\rangle \left|\alpha\right\rangle \left|\alpha\right\rangle \left|-\alpha\right\rangle \left|-\alpha\right\rangle \left|\alpha\right\rangle +\left|+\right\rangle \left|+\right\rangle \left|+\right\rangle \left|+\right\rangle \left|\alpha\right\rangle \left|\alpha\right\rangle \left|\alpha\right\rangle \left|\alpha\right\rangle \left|\alpha\right\rangle \right.\\
 & +\left.\left|-\right\rangle \left|+\right\rangle \left|+\right\rangle \left|-\right\rangle \left|-\alpha\right\rangle \left|-\alpha\right\rangle \left|\alpha\right\rangle \left|\alpha\right\rangle \left|-\alpha\right\rangle -\left|-\right\rangle \left|-\right\rangle \left|-\right\rangle \left|-\right\rangle \left|-\alpha\right\rangle \left|-\alpha\right\rangle \left|-\alpha\right\rangle \left|-\alpha\right\rangle \left|-\alpha\right\rangle \right).
\end{array}\label{eq:-7}
\end{equation}
At the last step, by performing the photon number measurement of the
coherent state in $a_{4}$ mode, which keeps $|\pm\alpha\rangle$
indistinguishable, they finally obtain the hybrid Omega-type MES 

\begin{equation}
\begin{array}{ll}
\left|\phi\right\rangle _{a_{3}bc_{3}d_{3}} & =\left|\Omega\right\rangle _{abcd},\end{array}\label{eq:-8}
\end{equation}
which is the desired concentrated hybrid Omega-type MES as shown in
Eq. (\ref{eq:omega}). 

Note that we have only used optical elements required for ECP of entangled
coherent state (\cite{Our_OECP} and references therein). This is
the advantage of hybrid ECPs that using solely one DOF and corresponding
technology the desired state can be concentrated. Therefore, the present
state can be concentrated using polarization DOF equivalently using
a polarization beamsplitter, half-waveplates, and the similar set
of auxiliary single- and two-qubit polarization states. Interestingly,
the success probability with polarization qubit based ECP is similar
to that obtained in the present case (discussed in detail in the next
section).

{There is non-zero probability of photon losses from both polarization and continuous variable part of the hybrid Omega-type MES during transmission and storage (see \cite{Near-deterministic_teleportation,Omkar} for detailed discussion). Further, imperfect sources, detectors or absorption in the rest of the optical elements used for various operations may also cause losses. The amplitude of coherent state reduces with photon loss rate parameter $\eta$ as $\alpha^{\prime}=\alpha\sqrt{1-\eta}$. Loss of single photons with polarization states will leave the state in continuous variable only. In general, loss of photons leads to mixedness of the reduced state, and in that case we require entanglement purification scheme for the distillation of entanglement. This scenario will be discussed in detail in the future.
}

\begin{figure}
\centering{}\includegraphics[scale=0.5]{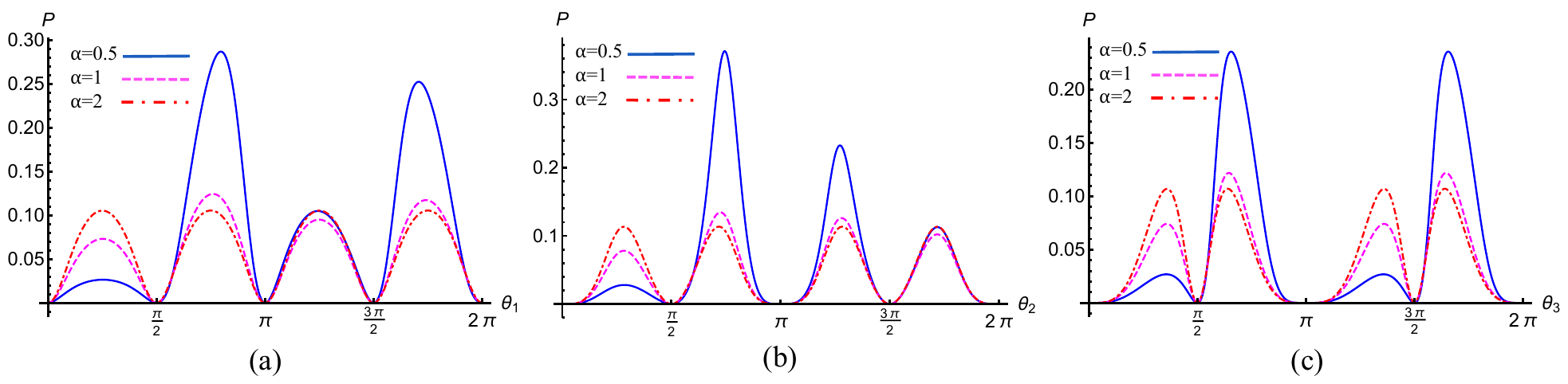}\caption{\label{fig:(Color-online)}(Color online) Variation of success probability
$(P)$ with (a) $\theta_{1}$, (b) $\theta_{2}$, and (c) $\theta_{3}$
considering the rest of the state parameters $\theta_{1}=\frac{\pi}{4}$,
$\theta_{2}=\frac{\pi}{4}$, and $\theta_{3}=\frac{3\pi}{8}$, wherever
needed. We have chosen the amplitude of coherent state $\alpha=0.5$
(smooth blue line), $\alpha=1$ (magenta dashed line), and $\alpha=2$
(red dot-dashed line). }
\end{figure}
\begin{figure}
\begin{centering}
\includegraphics[scale=0.5]{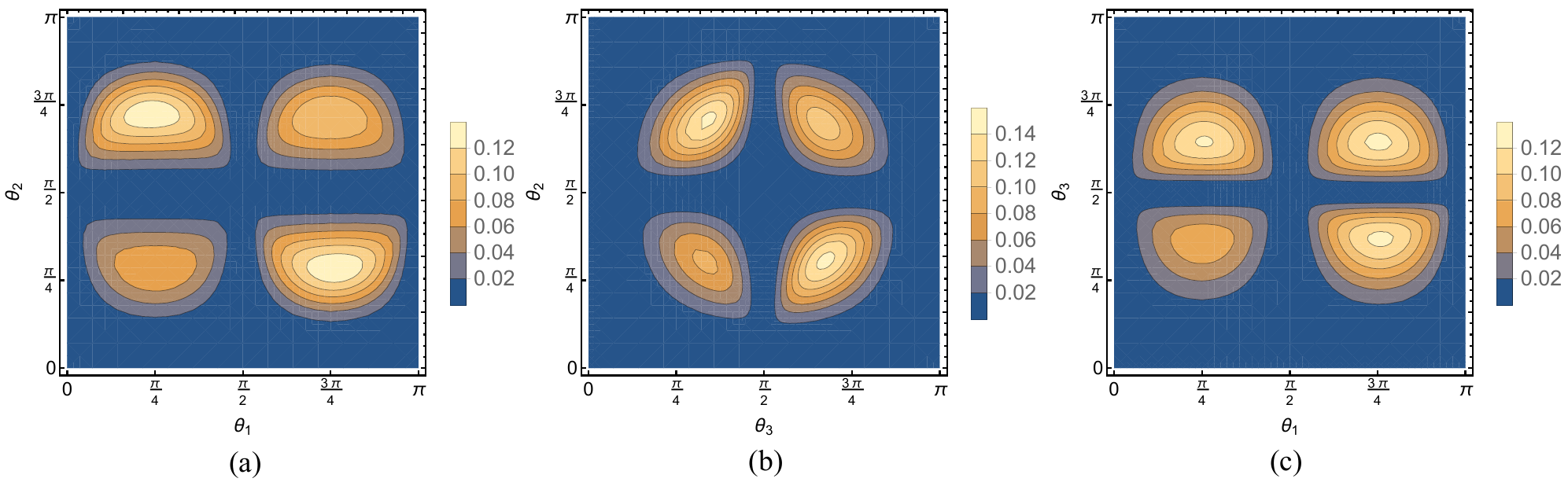}\caption{\label{fig:(Color-online)-(a)}(Color online) The contour plot shows
the variation of success probability $(P)$ with two state parameters
$\theta_{j}$. In all cases we assumed $\theta_{1}=\frac{\pi}{4}$,
$\theta_{2}=\frac{\pi}{4}$, and $\theta_{3}=\frac{3\pi}{8}$, wherever
needed, and the amplitude of coherent state $\alpha=1$. }
\par\end{centering}
\end{figure}

In general, we observed that the maxima (minima) of success probability
are obtained for state parameters $\theta_{j}=\frac{\left(2n+1\right)\pi}{4}$
$\left(\theta_{j}=\frac{n\pi}{2}\right)$ (cf. Fig. \ref{fig:(Color-online)-(a)}).
In Fig. \ref{fig:(Color-online)-(a)}, a symmetric variation of $P$
for $\alpha=1$ is illustrated. This shows that for a particular state
(with known amplitude of coherent state) the success probability depends
on the state parameters of non-MES. 

\section{Success probability\label{sec:Success-probability}}

In the previous section, we have obtained hybrid Omega-type MES from
the corresponding non-MESs with the success probability $P=4\left(N_{1}N_{2}\zeta\beta\delta\gamma\right)^{2},$
which can be easily calculated from Eq. (\ref{eq:-7}). Interestingly,
the same success probability for hybrid Cluster-type MES was recently
reported \cite{HES_Cluster_ECP}, which was of the order of that obtained
for CV Cluster state \cite{Our_OECP}. It is worth stressing here
using polarization DOF for concentration of the present Omega-type
HES analogously, the success probability is independent of $\alpha$
as the auxiliary state (\ref{eq:2_mode_ancilla}) will have $N_{1}=1$.
We have further discussed the variation of the success probability
$(P)$ with state parameters ($\zeta,\,\beta,\,\delta,$ and $\gamma$)
in Figs. \ref{fig:(Color-online)} and \ref{fig:(Color-online)-(a)}.
Exploiting the normalization condition $\zeta^{2}+\beta^{2}+\delta^{2}+\gamma^{2}=1$
for the real probability amplitudes, we can parameterize them as $\zeta=\cos\theta_{3}$,
$\beta=\sin\theta_{3}\cos\theta_{2}$, $\delta=\sin\theta_{3}\sin\theta_{2}\cos\theta_{1}$,
and $\gamma=\sin\theta_{3}\sin\theta_{2}\sin\theta_{1}$. In Fig.
\ref{fig:(Color-online)}, the maximum possible value of $P$ with
$\theta_{j}$ is obtained for $\alpha=0.5$ and further decreases
with increasing value of $\alpha$. The nature of the plot in Fig.
\ref{fig:(Color-online)} (b)-(c) is not as symmetric around $\theta_{j}=\frac{\left(2n+1\right)\pi}{4}$
as in Fig. \ref{fig:(Color-online)} (a). Clearly Fig. \ref{fig:(Color-online)}
shows that the dependence of $P$ on state parameters $\theta_{1}$,
$\theta_{2}$, and $\theta_{3}$ depends on the value of $\alpha$.
In brief, the symmetry and periodicity observed for $\alpha\geq1$
disappears for $\alpha<1$. 

\begin{figure}
\begin{centering}
\includegraphics[scale=0.7]{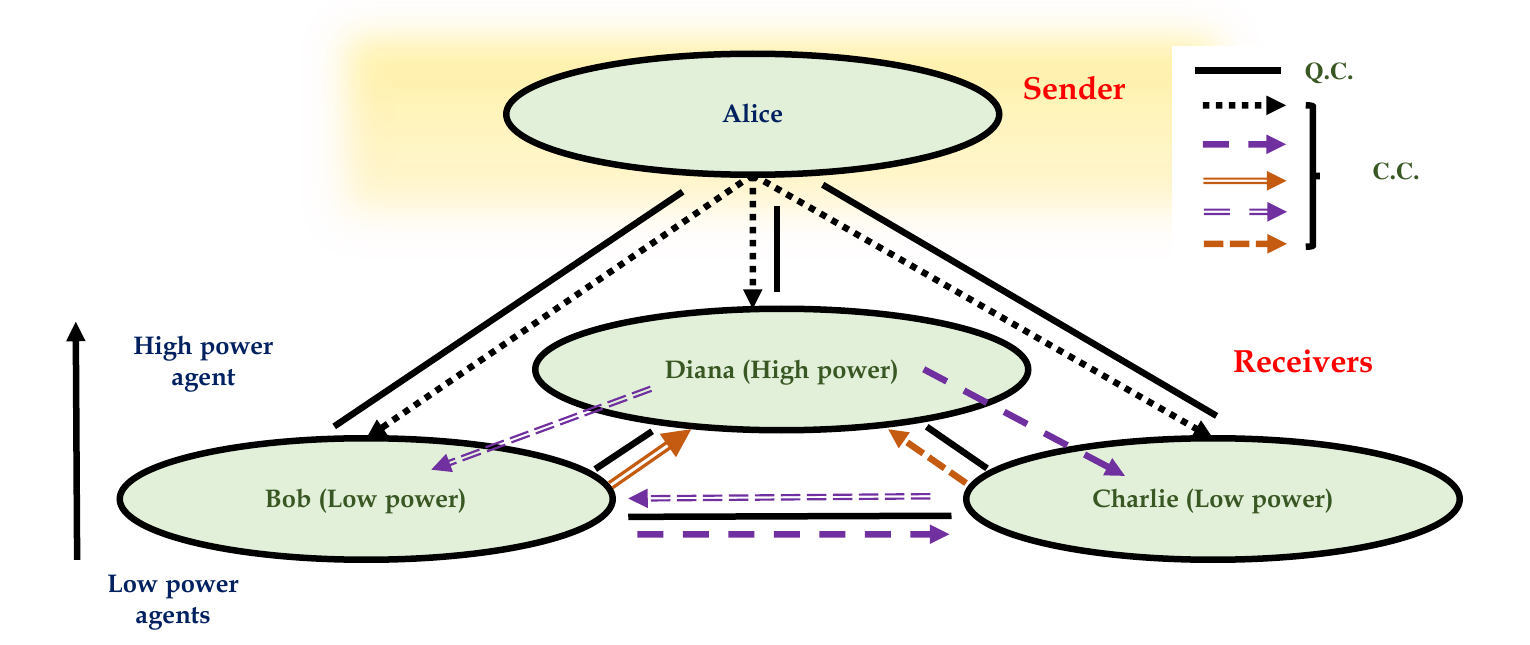}
\par\end{centering}
\caption{\label{fig:HQIS}(Color online) A schematic diagram for the proposed hierarchical communication network using hybrid entangled state.  The solid (black) line correspond to the hybrid entanglement shared among all the parties while arrows represent the classical communication required by each receiver to accomplish the task. Specifically, the dashed (black) arrow shows the information Alice has to broadcast to all the receivers. The dashed (purple) and double-dashed (purple) lines correspond to the classical communication required by Bob and Charlie, respectively. Diana requires information either from Bob (shown by double (brown) line) or Charlie (shown by dashed (brown) line).}
\end{figure}

\section{Application of Omega-type hybrid entangled state: Hierarchical quantum
information splitting of hybrid entanglement\label{sec:Application-of-Omega-type}}

In a hierarchical communication network of four participants, a boss
Alice (sender) distributes her quantum state among her agents (receivers)
Bob, Charlie, and Diana in such a way that receivers are arranged
in a hierarchy. Specifically, Diana (high power agent) can reconstruct
Alice's state with the cooperation of fewer agents (either Bob or
Charlie) after Alice's announcement; whereas Bob (Charlie), a low
power agent, can reconstruct Alice's state with the cooperation of
Diana as well as Charlie (Bob). Consequently, hierarchy exists among
the powers of agents (receivers) to reconstruct the quantum information
split between all the receivers. {A schematic diagram of the hierarchical communication network is shown in Fig. \ref{fig:HQIS}.} Suppose Alice wishes to teleport
(share) among her agents an unknown logical single-qubit $\left|\psi\right\rangle _{in}$
of the form 

\begin{equation}
\begin{array}{ccc}
\left|\psi\right\rangle _{in} & = & \left(\lambda\left|0_{L}\right\rangle +\eta\left|1_{L}\right\rangle \right)_{A_{0}},\end{array}\label{eq:input}
\end{equation}
where $\lambda$ and $\eta$ are complex numbers with $|\lambda|^{2}+|\eta|^{2}=1$,
and the subscript $L$ stands for the logical qubit $\left\{ \left|0_{L}\right\rangle =\left|+\right\rangle \left|\alpha\right\rangle ;\right.$
$\left.\left|1_{L}\right\rangle =\left|-\right\rangle \left|-\alpha\right\rangle \right\} $.
Notice that state (\ref{eq:input}) is CV-DV hybrid entangled state.
Alice chooses a hybrid Omega-type MES (\ref{eq:omega logic}) as quantum
channel 

\begin{equation}
\left|\Omega\right\rangle _{ch}=\frac{1}{2}\left[\left|0_{L}\right\rangle \left|0_{L}\right\rangle \left|0_{L}\right\rangle \left|0_{L}\right\rangle +\left|0_{L}\right\rangle \left|1_{L}\right\rangle \left|1_{L}\right\rangle \left|0_{L}\right\rangle +\left|1_{L}\right\rangle \left|0_{L}\right\rangle \left|0_{L}\right\rangle \left|1_{L}\right\rangle -\left|1_{L}\right\rangle \left|1_{L}\right\rangle \left|1_{L}\right\rangle \left|1_{L}\right\rangle \right]_{ABCD}\label{eq:channel}
\end{equation}
and shares the logical qubits $B$, $C$, and $D$ with her agents
Bob, Charlie, and Diana, respectively, keeping the logical qubit $A$
with herself. We know that hybrid entangled logical qubits undergo
decoherence during transmission transforming HES into non-MES, and
thus ECP discussed in the previous section is used by all the parties
to extract MES. The quantum channel can also be expressed as 

\begin{equation}
\begin{array}{ccc}
\left|\text{\ensuremath{\Omega}}\right\rangle _{ch} & = & \frac{1}{\sqrt{2}}\left[\left|0_{L}\right\rangle _{A}\left|\psi_{0_{L}}\right\rangle _{BCD}+\left|1_{L}\right\rangle _{A}\left|\psi_{1_{L}}\right\rangle _{BCD}\right],\end{array}\label{eq:3}
\end{equation}
where the combined states of Bob, Charlie and Diana are

\begin{equation}
\left|\psi_{0_{L}}\right\rangle _{BCD}=\frac{1}{\sqrt{2}}\left[\left|0_{L}\right\rangle \left|0_{L}\right\rangle \left|0_{L}\right\rangle +\left|1_{L}\right\rangle \left|1_{L}\right\rangle \left|0_{L}\right\rangle \right]_{BCD}\label{eq:4a}
\end{equation}
and
\begin{equation}
\left|\psi_{1_{L}}\right\rangle _{BCD}=\frac{1}{\sqrt{2}}\left[\left|0_{L}\right\rangle \left|0_{L}\right\rangle \left|1_{L}\right\rangle -\left|1_{L}\right\rangle \left|1_{L}\right\rangle \left|1_{L}\right\rangle \right]_{BCD}.\label{eq4a}
\end{equation}
The combined state of the system can be written as 

\begin{equation}
\left|\psi\right\rangle _{in}\otimes\left|\Omega\right\rangle _{ch}=\left(\lambda\left|0_{L}\right\rangle +\eta\left|1_{L}\right\rangle \right)_{A_{0}}\otimes\frac{1}{\sqrt{2}}\left[\left|0_{L}\right\rangle _{A}\left|\psi_{0_{L}}\right\rangle _{BCD}+\left|1_{L}\right\rangle _{A}\left|\psi_{1_{L}}\right\rangle _{BCD}\right],\label{eq:combined-1}
\end{equation}

Further, we define Bell-type state in the logical qubits as

\begin{equation}
\begin{array}{lclcc}
\left|\phi_{L}^{+}\right\rangle  & = & \frac{\left|0_{L}\right\rangle \left|0_{L}\right\rangle +\left|1_{L}\right\rangle \left|1_{L}\right\rangle }{\sqrt{2}} & = & \frac{\left|\phi^{+}\right\rangle \left|\phi_{+}\right\rangle +\left|\psi^{+}\right\rangle \left|\phi_{-}\right\rangle }{\sqrt{2}},\\
\left|\phi_{L}^{-}\right\rangle  & = & \frac{\left|0_{L}\right\rangle \left|0_{L}\right\rangle -\left|1_{L}\right\rangle \left|1_{L}\right\rangle }{\sqrt{2}} & = & \frac{\left|\phi^{+}\right\rangle \left|\phi_{+}\right\rangle -\left|\psi^{+}\right\rangle \left|\phi_{-}\right\rangle }{\sqrt{2}},\\
\left|\psi_{L}^{+}\right\rangle  & = & \frac{\left|0_{L}\right\rangle \left|1_{L}\right\rangle +\left|1_{L}\right\rangle \left|0_{L}\right\rangle }{\sqrt{2}} & = & \frac{\left|\psi^{+}\right\rangle \left|\psi_{+}\right\rangle +\left|\psi^{-}\right\rangle \left|\psi_{-}\right\rangle }{\sqrt{2}},\\
\left|\psi_{L}^{-}\right\rangle  & = & \frac{\left|0_{L}\right\rangle \left|1_{L}\right\rangle -\left|1_{L}\right\rangle \left|0_{L}\right\rangle }{\sqrt{2}} & = & \frac{\left|\psi^{+}\right\rangle \left|\psi_{+}\right\rangle -\left|\psi^{-}\right\rangle \left|\psi_{-}\right\rangle }{\sqrt{2}}.
\end{array}\label{eq:Logical_Bell1234}
\end{equation}

Here, the Bell states in polarization DOF are denoted as $\left|\phi^{\pm}\right\rangle =\frac{\left|0\right\rangle \left|0\right\rangle \pm\left|1\right\rangle \left|1\right\rangle }{\sqrt{2}}$
and $\left|\psi^{\pm}\right\rangle =\frac{\left|0\right\rangle \left|1\right\rangle \pm\left|1\right\rangle \left|0\right\rangle }{\sqrt{2}}$.
Also, the quasi-Bell states in coherent state DOF are $\left|\phi_{\pm}\right\rangle =N_{\pm}\left(\left|\alpha\right\rangle \left|\alpha\right\rangle \pm\left|-\alpha\right\rangle \left|-\alpha\right\rangle \right)$
and $\left|\psi_{\pm}\right\rangle =N_{\pm}\left(\left|\alpha\right\rangle \left|-\alpha\right\rangle \pm\left|-\alpha\right\rangle \left|\alpha\right\rangle \right)$,
with normalization constant for quasi-Bell state $N_{\pm}=\frac{1}{\sqrt{2\left(1\pm\exp\left[-4\mid\alpha\mid^{2}\right]\right)}}$.
This gives us a Bell-type basis in HES, and thus we can decompose
Alice's qubits $A_{0}A$ in Eq. (\ref{eq:combined-1}) into the above
logical Bell states (\ref{eq:Logical_Bell1234}) as follows

\begin{equation}
\begin{array}{lcl}
\left|\psi\right\rangle _{in}\otimes\left|\Omega\right\rangle _{ch} & = & \frac{1}{2}\left[\left|\phi_{L}^{+}\right\rangle _{A_{0}A}\left(\lambda\left|\psi_{0_{L}}\right\rangle _{BCD}+\eta\left|\psi_{1_{L}}\right\rangle _{BCD}\right)+\left|\phi_{L}^{-}\right\rangle _{A_{0}A}\left(\lambda\left|\psi_{0_{L}}\right\rangle _{BCD}-\eta\left|\psi_{1_{L}}\right\rangle _{BCD}\right)\right.\\
 & + & \left.\left|\psi_{L}^{+}\right\rangle _{A_{0}A}\left(\lambda\left|\psi_{1_{L}}\right\rangle _{BCD}+\eta\left|\psi_{0_{L}}\right\rangle _{BCD}\right)+\left|\psi_{L}^{-}\right\rangle _{A_{0}A}\left(\lambda\left|\psi_{1_{L}}\right\rangle _{BCD}-\eta\left|\psi_{0_{L}}\right\rangle _{BCD}\right)\right].
\end{array}\label{eq:combined in bell-1-1}
\end{equation}

This shows that after Alice performs a logical Bell (basis) measurement
on her qubits $A_{0}A$ and broadcasts her outcome the reduced state
is shared among Bob, Charlie and Diana. Alice's measurement outcomes
and the reduced tripartite state sharing among the receivers is summarized
in Table \ref{tab:According-to-Eq.}.

\begin{table}[H]
\begin{centering}
\begin{tabular}{cc}
\toprule 
Alice's logical Bell measurement outcome & Combined state of all three agents\tabularnewline
\midrule
$\left|\phi_{_{L}}^{\pm}\right\rangle _{A_{0}A}$ & $\left(\lambda\left|\psi_{0_{L}}\right\rangle \pm\eta\left|\psi_{1_{L}}\right\rangle \right)_{BCD}$\tabularnewline
$\left|\psi_{_{L}}^{\pm}\right\rangle _{A_{0}A}$ & $\left(\lambda\left|\psi_{1_{L}}\right\rangle \pm\eta\left|\psi_{0_{L}}\right\rangle \right)_{BCD}$\tabularnewline
\bottomrule
\end{tabular}
\par\end{centering}
\caption{\label{tab:According-to-Eq.}According to Eq. (\ref{eq:combined in bell-1-1}),
Alice's logical Bell measurement outcomes shown in Column 1 and the
corresponding combined states of all the three agents are shown in
Column 2.}
\end{table}
For instance, if Alice's measurement outcome is $\left|\phi_{_{L}}^{+}\right\rangle _{A_{0}A}$
then the corresponding receivers' state collapses to $\left(\lambda\left|\psi_{0_{L}}\right\rangle +\eta\left|\psi_{1_{L}}\right\rangle \right)_{BCD}$,
which can further be expanded using the values of $\left|\psi_{0_{L}}\right\rangle _{BCD}$
and $\left|\psi_{1_{L}}\right\rangle _{BCD}$ from Eqs. (\ref{eq:4a}-\ref{eq4a})
as follows

\begin{equation}
\left|\psi^{\prime}\right\rangle =\frac{1}{\sqrt{2}}\left[\lambda\left(\left|0_{L}\right\rangle \left|0_{L}\right\rangle \left|0_{L}\right\rangle +\left|1_{L}\right\rangle \left|1_{L}\right\rangle \left|0_{L}\right\rangle \right)+\eta\left(\left|0_{L}\right\rangle \left|0_{L}\right\rangle \left|1_{L}\right\rangle -\left|1_{L}\right\rangle \left|1_{L}\right\rangle \left|1_{L}\right\rangle \right)\right]_{BCD}.\label{eq:bob}
\end{equation}
Notice that Alice's quantum information is split among the agents,
and now $\left|\psi\right\rangle _{in}$ has to be hierarchically
recovered by agents. In what follows, we consider two cases, i.e.,
when the higher and lower power agents recover the quantum information.

\textbf{Case 1:} If the agents decide that Diana (high power agent)
will recover the quantum state sent by Alice, then we can decompose
Eq. (\ref{eq:bob}) as

\begin{equation}
\left|\psi^{\prime}\right\rangle =\frac{1}{\sqrt{2}}\left[\left|0_{L}\right\rangle _{B}\left|0_{L}\right\rangle _{C}\left(\lambda\left|0_{L}\right\rangle +\eta\left|1_{L}\right\rangle \right)_{D}+\left|1_{L}\right\rangle _{B}\left|1_{L}\right\rangle _{C}\left(\lambda\left|0_{L}\right\rangle -\eta\left|1_{L}\right\rangle \right)_{D}\right].\label{eq:Diana_Recovers}
\end{equation}

It is to be noted that both Bob and Charlie measure in the logical
qubit basis $\left\{ \left|0_{L}\right\rangle ,\left|1_{L}\right\rangle \right\} $
and would always have the same measurement outcomes as shown in Column
1 of Table \ref{tab:According-to-Eq.-1}, hence the communication
from one of them in addition to Alice's announcement would be enough
to enable Diana to reconstruct the unknown state sent by Alice. The
significance of using this logical Omega-type HES lies in the fact
that $Z$ gate to be applied by Diana can be applied using waveplate
on single qubit polarization states. Note that $Z$ gate applied on
coherent state gives errors due to nonorthogonality of $|\pm\alpha\rangle$
\cite{Near-deterministic_teleportation}, and thus single qubit in
HES provides that advantage.

\begin{table}[H]
\begin{centering}
\begin{tabular}{ccc}
\toprule 
Measurement outcome of Bob and Charlie & Diana's state before operation  & Diana's operation\tabularnewline
\midrule
$\left|0_{L}\right\rangle _{B}\left|0_{L}\right\rangle _{C}$ & $\left(\lambda\left|0_{L}\right\rangle +\eta\left|1_{L}\right\rangle \right)_{D}$ & $I$\tabularnewline
$\left|1_{L}\right\rangle _{B}\left|1_{L}\right\rangle _{C}$ & $\left(\lambda\left|0_{L}\right\rangle -\eta\left|1_{L}\right\rangle \right)_{D}$ & $Z$\tabularnewline
\bottomrule
\end{tabular}
\par\end{centering}
\caption{\label{tab:According-to-Eq.-1} Bob's and Charlie's measurement outcomes,
Diana's collapsed state, and the unitary operations to be applied
by Diana to recover the unknown state sent by Alice. }
\end{table}

\textbf{Case 2:} If the agents decide that Bob (low power agent) will
recover the quantum state sent by Alice, then we can decompose Eq.
(\ref{eq:bob}) as 

\begin{equation}
\begin{array}{ccc}
\left|\psi^{\prime}\right\rangle  & = & \frac{1}{2}\left[\left|\phi_{L}^{+}\right\rangle _{CD}\left(\lambda\left|0_{L}\right\rangle -\eta\left|1_{L}\right\rangle \right)_{B}+\left|\phi_{L}^{-}\right\rangle _{CD}\left(\lambda\left|0_{L}\right\rangle +\eta\left|1_{L}\right\rangle \right)_{B}\right.\\
 & + & \left.\left|\psi_{L}^{+}\right\rangle _{CD}\left(\lambda\left|1_{L}\right\rangle +\eta\left|0_{L}\right\rangle \right)_{B}+\left|\psi_{L}^{-}\right\rangle _{CD}\left(\eta\left|0_{L}\right\rangle -\lambda\left|1_{L}\right\rangle \right)_{B}\right].
\end{array}\label{eq:Bob_Recovers}
\end{equation}

Here, Charlie and Diana are required to perform a joint logical Bell
(basis) measurement $\left\{ \left|\phi_{L}^{\pm}\right\rangle ,\left|\psi_{L}^{\pm}\right\rangle \right\} $
and it is therefore necessary for Bob seeking the cooperation of both
of them and Alice to reconstruct the unknown state sent by Alice (summarized
in Table \ref{tab:According-to-Eq.-2}). Notice that Bob's ignorance
is 2 bits, and Bob's state is a mixed state $\mathbb{I}_{2}^{L}$
in logical qubits after tracing out Charlie's and Diana's logical
qubits. Bob (Charlie) too can apply Pauli operations on single qubit
polarization states to reconstruct the state instead of operations
on coherent states.

It is worth mentioning here that if Charlie decides to reconstruct
the state, this case would be same as Case 2. Similarly, for the rest
of the three cases of Alice's measurement outcomes $\left|\phi_{_{L}}^{-}\right\rangle _{A_{0}A}$
and $\left|\psi_{_{L}}^{\pm}\right\rangle _{A_{0}A}$ as shown in
Table \ref{tab:According-to-Eq.}, the whole scheme for both high
and low power agents will be followed exactly the same way as we have
shown for $\left|\phi_{_{L}}^{+}\right\rangle _{A_{0}A}$. 

\begin{table}[h]
\begin{centering}
\begin{tabular}{ccc}
\toprule 
Measurement outcome of Charlie and Diana  & Bob's state before operation  & Bob's operation\tabularnewline
\midrule
$\left|\phi_{_{L}}^{+}\right\rangle _{CD}$ & $\left(\lambda\left|0_{L}\right\rangle -\eta\left|1_{L}\right\rangle \right)_{B}$ & $Z$\tabularnewline
$\left|\phi_{_{L}}^{-}\right\rangle _{CD}$ & $\left(\lambda\left|0_{L}\right\rangle +\eta\left|1_{L}\right\rangle \right)_{B}$ & $I$\tabularnewline
$\left|\psi_{_{L}}^{+}\right\rangle _{CD}$ & $\left(\lambda\left|1_{L}\right\rangle +\eta\left|0_{L}\right\rangle \right)_{B}$ & $X$\tabularnewline
$\left|\psi_{_{L}}^{-}\right\rangle _{CD}$ & $\left(\eta\left|0_{L}\right\rangle -\lambda\left|1_{L}\right\rangle \right)_{B}$ & $iY$\tabularnewline
\bottomrule
\end{tabular}
\par\end{centering}
\caption{\label{tab:According-to-Eq.-2} Charlie's and Diana's measurement
outcomes, Bob's collapsed state, and the unitary operations applied
by Bob to recover the unknown state sent by Alice.}
\end{table}

Notice that while recovering the initial state after Alice's measurement
outcome, Bob/Charlie (Diana) require(s) the help from both Diana and
Charlie/Bob (either Bob or Charlie), and clearly Diana has higher
power than Bob (or Charlie) to recover Alice's state. Thus, a hierarchy
exists among the agents in such a way that Bob requires more information
than that by Diana for the reconstruction of the unknown quantum state
sent by Alice. This is why this scheme is referred to as HQIS. Similar
HQIS of HES can be proposed using the Cluster-type HES \cite{HES_Cluster_ECP}.
Further, by modifying our hybrid HQIS scheme proposed here, we aim
to investigate the hybrid hierarchical quantum secret sharing scheme
in our future work. 

Before concluding the work it is imperative to stress that the present
HES cannot be used for HQIS of an arbitrary two-qubit state as according
to Nielsen's criterion \cite{Nielsen_Criterion} the present state
(Omega-type HES) cannot be transformed to the product of two Omega
states (hyper entangled state in the present case) using local operation
and classical communication \cite{Channels_for_QT}. Therefore, we
restricted ourselves to the application of hybrid Omega-type MES in
HQIS of HES.

\section{Conclusion\label{sec:Conclusion}}

As emphasized in the introduction, the success of a quantum communication
or computation task, and quantum technology in general, MES is an
essential quantum resource. First of all, it is difficult to distribute
a MES between distant parties and even if we succeed in achieving
that, it would be hard to maintain entanglement for a longer period
of time. Thus, a MES would be transformed into a non-MES due to its
interaction with the surroundings before its application. Therefore,
it becomes necessary to extract the MES from the non-MESs using an
ECP before performing an application using such quantum resources.
Essentially, ECPs/EPPs are the initial steps for performing entangled
state based long-distance quantum communication and secure computation. 

Here, we have chosen a hybrid Omega-type MES inspired by the possibility
of its application in quantum communication over network. For the
sake of completeness of the present work and to achieve the unit success
probability, first we have proposed ECP of the HES using a single-
and two-mode superposition of coherent states. Interestingly, the
proposed ECP can be performed with linear optics and available technology
only. The proposed scheme not only provides ECP but a possibility
of amplification of the transmitted modes, which may have advantages
in some application, for instance, to counter photon loss in long distance
quantum communication. Even when this amplification is not required
the desired state can be obtained by performing a photon number measurement
on one of the modes after beamsplitter as it would not disturb the
entanglement. We further obtained the success probability $P$ for
the ECP and found that it depends not only on the amplitude of coherent
state $(\alpha)$ but the coefficients of the non-MES. 

Further, to accomplish our aim, we have discussed an application of
the obtained hybrid Omega-type MES in quantum information splitting
over a hierarchical network. Specifically, we have performed HQIS
of a HES using the present scheme, which is the first such hierarchical
communication scheme. The advantage of the present HES is visible
here as the receiver can recover the shared quantum information by
performing single qubit polarization rotation, which can be implemented
with less errors than the corresponding operations on coherent state.
The present HQIS scheme can also be viewed as an application of hybrid
Omega-type MES in a hierarchical controlled quantum teleportation
network in view of some of the recent schemes along the same line
\cite{Ba_An_Hierarchically_controlling_QT}. Similarly, the proposed
ECP and the concerned state can also be used in HES based two- and
multi-party networks for secure quantum communication and computation
\cite{QD,AQD,QC}. Thus, hybrid entanglement is a powerful resource
for such communication tasks, and ECPs for such states are the demand
of time as we are generally going to implement communication tasks
over quantum networks. We conclude this work focused on the first
hybrid quantum information splitting over hierarchical network using
Omega-type HES and an ECP required to accomplish the task efficiently.
with a hope that it can be realized experimentally and would be significant
in long-range quantum communication and distributed quantum computing
in the near future.

\noindent \textbf{ACKNOWLEDGEMENTS}

\noindent KT acknowledges the financial support from the Operational Programme
Research, Development and Education - European Regional Development
Fund project no. CZ.02.1.01/0.0/0.0/16\_019/0000754 of the Ministry
of Education, Youth and Sports of the Czech Republic. PM thanks Peng
Cheng Laboratory (PCL), Shenzhen, China for the hospitality provided
during her visit. Authors also thank Prof. Anirban Pathak for his
interest in the work and helpful suggestions.

\end{document}